\documentclass[12pt]{article}

\usepackage{hyperref}
\usepackage{amssymb}
\usepackage{mathrsfs}
\usepackage{amsthm}
\usepackage{amsmath}
\usepackage{graphicx}
\usepackage{xcolor}

\begin{document}

\title{\Large\bf Trapped string states in  AdS$_5$ black hole geometry:\\ A path toward Hawking radiation}

\author{{\sc Harpreet Singh}\thanks{\tt harpreetsingh@iitg.ac.in} \ and {\sc Malay K. Nandy}\thanks{\tt mknandy@iitg.ac.in}\\
{\it Department of Physics, Indian Institute of Technology Guwahati}\\
{\it Guwahati 781 039, India}
}

\date{}

\maketitle

\begin{abstract}

We investigate the quantum dynamics of a closed bosonic string in the curved spacetime of an AdS$_5$-Schwarzschild black hole. Starting from the Polyakov action, we perform a canonical quantization of the string and formulate its quantum mechanical equation of motion in the Schr\"odinger (string coordinate) representation.  This framework facilitates in obtaining quantum mechanical wave equation governing the radial and angular modes of the string.  A central result of our analysis is the emergence of a trapping radius in the exterior region of the black hole. Near this radius, the radial motion of the string is governed by an effective potential that supports small, quantized oscillations, akin to a quantum harmonic oscillator. This behavior indicates a localization of the string at the trapping surface, where it becomes dynamically confined. The angular sector of the wave function is found to be governed by the confluent Heun equation, yielding confluent Heun functions as the angular part of the wave function. The emergence of a trapping surface is analogous to the stretched horizon proposed by Susskind in the context of black hole complementarity. The quantized harmonic oscillation of the string at the trapping radius complements with the Planck's black body whence the string can emit black-body radiation. Thus, the quantum dynamics of strings in black hole spacetimes offers a novel path to probing the quantum origin of Hawking radiation.

\end{abstract}

\newpage

\tableofcontents

\section{Introduction}

The behavior of matter near the event horizon of a black hole remains a central question in theoretical physics. Quantum mechanical effects in this regime are not only of fundamental interest but are also pivotal for addressing some of the deepest challenges in modern physics, notably the microscopic origin of black hole entropy and the information loss paradox. Understanding the quantum dynamics of matter as it approaches the horizon may thus offer critical insights into the interplay between gravity and quantum mechanics.

String theory, as a leading candidate for a unified theory of quantum gravity, provides a natural framework for exploring these phenomena. Strings, being extended objects, offer a richer set of degrees of freedom than point particles, and are therefore particularly well-suited for probing the physics of near-horizon geometries.

The dynamics of particles and strings in curved spacetimes are of considerable interest, not only from a theoretical standpoint but also for their implications in astrophysics. These dynamics play a significant role in phenomena such as the structure and stability of accretion disks, gravitational lensing, and the emission of gravitational waves. Consequently, their study has broad relevance across both fundamental physics and observational astronomy \cite{PhysRevD.93.084012, PhysRevD.83.124015, Panis2019, PhysRevD.95.084037, PhysRevD.83.044053, PhysRevD.99.044012, Yi_2020, PhysRevD.50.R618, doi:10.1142/S0217732307022815, Hosur2016, Gur-Ari2016, PhysRevD.95.066014, PhysRevD.89.086011, Ma2020, Ma2022, Maldacena2016, PandoZayas2010}.

Various studies have explored the motion of particles in black hole spacetimes. For instance, Panis et al. \cite{Panis2019} showed that a particle exhibits periodic motion in the Schwarzschild background, whereas motion in Kerr-Newman spacetime tends to be quasi-periodic. In contrast, the motion of charged particles in magnetized or charged black holes is often chaotic \cite{PhysRevD.95.084037, PhysRevD.83.044053, PhysRevD.99.044012, Yi_2020, PhysRevD.50.R618, doi:10.1142/S0217732307022815}.

The dynamics of strings near black hole horizons have similarly been shown to exhibit chaotic behavior \cite{PandoZayas2010, PhysRevD.89.086011, Ma2020, Ma2022, LI2025139164}. Pando Zayas and Terrero-Escalante \cite{PandoZayas2010} studied strings in the AdS$_5$-Schwarzschild background and, using power spectra and Poincar\'e sections, demonstrated the occurrence of chaos via a large Lyapunov exponent. Ma et al. \cite{PhysRevD.89.086011} investigated string dynamics in the AdS-Gauss-Bonnet background and identified a critical value of the Gauss-Bonnet parameter beyond which the system transitions from regular to chaotic behavior.

In further studies, the effects of Hawking temperature and the Lifshitz dynamical exponent on string dynamics were analyzed in charged black brane spacetimes \cite{Ma2020}. It was found that below a critical Lifshitz exponent, strings are captured by the brane, while above it, they can escape even when initially placed near the horizon. Quasi-periodic motion was observed when strings were placed far from the black brane. The impact of the winding number and hyperscaling violation exponent on the onset of chaos was also explored. Related investigations extended this analysis to conformal black holes \cite{Ma2022} and charged Kiselev black holes \cite{LI2025139164}.

The quantization of strings in curved backgrounds was addressed in early works such as de Vega and S\'anchez \cite{DEVEGA1987320}, who solved the equations of motion and constraints by treating the spacetime geometry exactly and string excitations perturbatively. They derived the mass spectrum and vertex operators in de Sitter space. Semiclassical quantization of oscillatory circular strings in de Sitter and anti-de Sitter backgrounds was carried out in \cite{PhysRevD.51.6917}, where it was shown that mass levels in AdS remain evenly spaced, while those in de Sitter decay via tunneling.

The dynamics of the Nambu–Goto string in anti-de Sitter (AdS) spacetime have been extensively investigated in the context of the AdS/CFT correspondence, particularly in Refs. \cite{Ishii2015, Ishii2016, PhysRevD.98.086007}. These studies are motivated by the foundational idea of the AdS/CFT duality proposed by Maldacena \cite{Maldacena1999}, which posits an equivalence between a gravitational theory in AdS space and a conformal field theory (CFT) defined on its boundary.

Classical string propagation in and near black hole horizons has also been studied extensively. Ref. \cite{PhysRevD.53.3296} examined the behavior of strings near the singularity and horizon, finding divergent momentum and size at the singularity, with radiation-like behavior emerging in the near-singularity regime. Other studies have explored string motion in the Kerr-Newman background \cite{FROLOV1989255}, demonstrating that time-dependent configurations can be mapped to geodesic motion in an effective three-dimensional metric. Infinitely long stationary string solutions were obtained outside the event horizon.

String instabilities have also been identified in Schwarzschild and Reissner-Nordström spacetimes \cite{PhysRevD.47.4498}, with stable time components but unstable radial components under small fluctuations. The behavior of ring-shaped strings in FRW and Schwarzschild backgrounds was explored in \cite{PhysRevD.49.763}, showing that initial conditions determine whether a string is absorbed or escapes. In \cite{PhysRevD.51.6929}, elliptic function solutions for open and closed strings in curved spacetimes were derived, and it was shown that straight strings falling into Schwarzschild black holes grow unboundedly near the singularity.

In Ref. \cite{PhysRevD.49.763}, the dynamics of strings in both cosmological and black hole spacetimes were studied using a ring-shaped ansatz. By solving the equations of motion and constraint equations, the authors analyzed the evolution of string configurations in different backgrounds. In the case of a Friedmann-Lema\^itre-Robertson-Walker (FLRW) universe, it was shown that following the Big Bang, the string's size initially grows while its energy decreases, eventually reaching a stable oscillatory regime characterized by constant size and energy. In contrast, for the Schwarzschild black hole background, the fate of the string---whether it is absorbed by the black hole or escapes---was found to depend sensitively on its initial conditions.

Further work on classical string dynamics in curved spacetimes was conducted in Ref. \cite{PhysRevD.51.6929}, where solutions for open and closed strings, as well as for finitely and infinitely long strings and multistring configurations, were constructed in terms of elliptic functions. In the Schwarzschild background, it was demonstrated that a straight string falling non-radially toward the singularity undergoes unbounded growth in length as it approaches the singularity. Notably, no multistring solutions were found to exist within the interior region of the Schwarzschild black hole.

Circular strings in Schwarzschild, Reissner-Nordström, and de Sitter spacetimes were investigated in \cite{PhysRevD.50.2623}, revealing divergent radial perturbations as $r \to 0$, while angular perturbations remain bounded.

A particularly intriguing idea is the conjecture that strings can spread over the horizon, forming a membrane-like structure known as the ``stretched horizon'' \cite{membranebook, THOOFT1985727, THOOFT1990138, GtHooft_1991, PhysRevLett.71.2367, PhysRevD.48.3743, PhysRevD.49.6606, PhysRevD.50.2725, PhysRevD.60.024012}. 't Hooft \cite{THOOFT1985727, THOOFT1990138} proposed that the black hole horizon may be governed by an operator algebra on a two-dimensional surface, akin to the worldsheet description of strings. While the black hole is treated as a classical statistical system on its membrane (horizon), the string remains a quantum object, and the two descriptions may be connected via Wick rotation \cite{GtHooft_1991}.

Susskind advanced this idea by proposing a complementary picture: a string falling into a black hole appears to pass smoothly through the horizon to an infalling observer, but from the perspective of an external observer, it spreads over the stretched horizon, carrying its information with it \cite{PhysRevLett.71.2367}. He further suggested that the stretched horizon can absorb and emit radiation, and that these macroscopic behaviors arise from coarse-graining microscopic quantum degrees of freedom \cite{PhysRevD.48.3743, PhysRevD.49.6606}. Thermalization occurs through the branching and diffusion of string bits, which are eventually re-emitted as Hawking radiation \cite{PhysRevD.50.2725}.

In exact Schwarzschild geometry, it was shown that strings spread over the angular directions of the horizon while radial spreading is suppressed due to Lorentz contraction \cite{PhysRevD.60.024012}. Additionally, it has been proposed that near the black hole horizon, strings may undergo a Hagedorn transition \cite{PhysRevD.51.1793}, reflecting a limiting temperature of string theory.

Current-carrying strings have also been studied in various black hole backgrounds \cite{ALarsen_1993, LARSEN1991375, ALarsen_1994, AndreiVFrolov_1999, PhysRevD.79.065029}. In particular, relativistic strings with current moving along the axis of a Kerr black hole were analyzed in \cite{PhysRevD.79.065029}, with energy and escape conditions derived. Chaotic behavior was observed for strings in Schwarzschild spacetimes \cite{AndreiVFrolov_1999}, with a critical energy threshold above which chaos emerges. In \cite{ALarsen_1994}, microscopic current-carrying strings displayed chaotic motion around the equator of Schwarzschild black holes, including transmutation and capture scenarios.

Stable configurations of charged strings in Kerr-Newman spacetimes were analyzed in \cite{LARSEN1991375}, while a general Hamiltonian framework for current-carrying strings was developed in \cite{ALarsen_1993}, demonstrating that the dynamics can, in some cases, be effectively reduced to a point particle system.

In this work, we focus upon the dynamics of a closed bosonic string in the background of an AdS$_5$ Schwarzschild black hole. Using the Polyakov action, we canonically quantize the string and derive a Schr\"odinger-type equation governing its dynamical behavior. Our key finding is the emergence of a trapping radius in the exterior region of the black hole, where the string becomes localized and undergoes quantized oscillatory motion akin to that of a harmonic oscillator. This phenomenon is strikingly similar to the concept of the stretched horizon, introduced by Susskind, which posits a membrane-like surface outside the event horizon. Additionally, we show that the angular part of the string wave function is governed by the confluent Heun equation, having confluent Heun function as its solution. 

Furthermore, the quantized harmonic oscillations of the string at the trapping surface allow it to behave similarly to a Planckian black-body radiator. This correspondence offers a compelling microscopic picture of how Hawking radiation might emerge from string degrees of freedom. In particular, it suggests a mechanism wherein the thermal emission from black holes can be understood as arising from the quantized vibrations of strings localized at the trapping surface.

The remainder of the paper is organized as follows. In Section \ref{ads5}, we lay out the Lagrangian formulation of string dynamics from the Polyakov action in the AdS$_5$-Schwarzschild background. Section \ref{dyn} presents the Hamiltonian formulation, while in Section \ref{qm}, we perform canonical quantization and obtain the corresponding quantum mechanical equation in the Schr\"odinger (string coordinate) representation. In Section \ref{trap}, we analyze the radial equation and identify the emergence of a trapping radius. The angular sector is examined in Section \ref{ang}, where the angular part of the wave function is solved for. Finally, Section \ref{concl} offers a summary of the results with concluding remarks.

\section{String in ${\bf AdS}_5$ background}
\label{ads5}

The dynamics of a string in the background of a curved spacetime is described by the Polyakov action \cite{POLYAKOV1981207, Green_Schwarz_Witten_2012, Polchinski_1998, Zwiebach_2009, Becker_Becker_Schwarz_2006}, 
\begin{equation}
 S=-\frac{1}{4 \pi \alpha^{\prime}}\int_M d\tau d\sigma \sqrt{-h} h^{ab} g_{\mu\nu}\partial_a X^\mu \partial_b X^\nu, 
 \label{PolyakovS}
\end{equation}
where $X^\mu(\tau , \sigma)$ is the  string worldsheet coordinate with respect to the background spacetime characterized by the metric $g_{\mu\nu}$. The metric $h_{ab}$ characterizes the two-dimensional  worldsheet manifold $M$, with the indices $(a,b) \in \{\tau, \sigma\}$. The Regge slope $\alpha^{\prime}$ has its origin in the  string tension, $T=(2 \pi \alpha^{\prime})^{-1}$.

The two dimensional string worldsheet, along with having two reparametrization gauge invariances, has an additional symmetry with respect to Weyl (conformal) rescaling leaving $\sqrt{-h}h^{ab}$ invariant. This makes it possible to choose a gauge in which $h_{ab}$ is diagonal,  with $h_{ab}={\rm diag}(-1,+1)$.

We consider the dynamics of the string in the background of a five-dimensional AdS-Schwarzschild spacetime,  with its metric given by 
\begin{align}
 ds^2=-f(r) dt^2+\frac{dr^2}{f(r)}+r^2(d\theta^2+\sin^2\theta d\chi^2+\cos^2\theta d\phi^2),
 \label{metric}
\end{align}
so that $g_{\mu\nu}={\rm diag}(-f,f^{-1},r^2, r^2 \sin^2\theta, r^2 \cos^2\theta )$, with $X^{\mu}=(t,r,\theta,\chi, \phi)$.

We shall study the dynamics of a closed  string  with the embedding 
\begin{align}
 t=t(\tau),\  r=r(\tau),\ \theta=\theta(\tau), \ \phi=\phi(\tau), \ {\rm and}\ \chi=k \sigma,
 \label{emb}
\end{align}
with $k$ a constant. With this embedding, the string Lagrangian obtained from the Polyakov action \ref{PolyakovS} turns out to be    
\begin{align}
 \mathcal{L}= -\frac{1}{4\pi \alpha^{\prime}}\left[f\dot t^2-\frac{\dot r^2}{f}-r^2(\dot \theta^2+ \dot \phi^2 \cos^2 \theta)+ k^2 r^2  \sin^2 \theta\right],
\end{align}
in the AdS$_5$ background described by the metric \ref{metric}. Here and elsewhere an overdot represents  derivative with respect to $\tau$.

Varying the action \ref{PolyakovS} with respect to $h_{ab}$ gives rise to the  nontrivial constraints \cite{Polchinski_1998}
\begin{align}
 g_{\mu\nu}\left(\frac{\partial X^{\mu}}{\partial \tau}  \frac{\partial X^{\nu}}{\partial \tau}+ \frac{\partial X^{\mu}}{\partial \sigma}  \frac{\partial X^{\nu}}{\partial \sigma}\right)=0
 \label{con}
\end{align}
and 
\begin{align}
 g_{\mu\nu}\frac{\partial X^{\mu}}{\partial \tau}  \frac{\partial X^{\nu}}{\partial \sigma}=0.
 \label{con2}
\end{align}

Employing the metric \ref{metric}, the first constraint equation \ref{con}  yields 
\begin{align}
-f\dot t^2+\frac{\dot r^2}{f}+r^2(\dot \theta^2+ \dot \phi^2 \cos^2 \theta)+ k^2 r^2  \sin^2 \theta=0,
\label{con7}
\end{align}
with the embedding \ref{emb}, whereas the second constraint \ref{con2} is identically satisfied.

\section{Hamiltonian dynamics of the string}
\label{dyn}

The canonical momenta conjugate to the coordinates $t$, $r$, $\theta$ and $\phi$ are given by
\begin{align}
 p_t=\frac{\partial \mathcal{L}}{\partial \dot t}=-\frac{1}{2 \pi \alpha^{\prime}} f \dot t,
\end{align}
\begin{align}
 p_r=\frac{\partial \mathcal{L}}{\partial \dot r}=\frac{1}{2 \pi \alpha^{\prime}}\frac{\dot r}{f},
\end{align}
\begin{align}
 p_{\theta}=\frac{\partial \mathcal{L}}{\partial \dot \theta}=\frac{1}{2 \pi \alpha^{\prime}}r^2 \dot \theta,
\end{align}
and
\begin{align}
 p_{\phi}=\frac{\partial \mathcal{L}}{\partial \dot \phi}=\frac{1}{2 \pi \alpha^{\prime}} r^2 \dot \phi \cos^2 \theta. 
\end{align}
Hence, the Hamiltonian $\mathcal{H}=p_t\dot t+p_r \dot r+p_{\theta} \dot \theta+p_{\phi} \dot \phi-\mathcal{L}$ turns out to be 
\begin{align}
 \mathcal{H}=\pi \alpha^{\prime}\left( f p_r^2 +\frac{p_{\theta}^2}{r^2}+\frac{p_{\phi}^2}{r^2 \cos^2 \theta}-\frac{p_t^2}{f}\right)+\frac{1}{4\pi \alpha^{\prime}}k^2 r^2  \sin^2 \theta.
 \label{ham}
\end{align}

Since the Hamiltonian \ref{ham} is cyclic in the coordinates $t$ and $\phi$, the corresponding conjugate momenta $p_t$ and $p_\phi$ are integrals of motion. Hamilton's equations  for $t$ and $\phi$ are expressed as 
\begin{align}
 \dot t = \frac{\partial \mathcal{H}}{\partial p_t}=- \frac{2 \pi \alpha^{\prime} P_0}{f(r)}
 \label{m1}
\end{align}
and 
\begin{align}
 \dot \phi = \frac{\partial \mathcal{H}}{\partial p_\phi}= \frac{2 \pi \alpha^{\prime} P_4}{r^2\cos^2\theta},
 \label{m4}
\end{align}
where $P_0=p_t=$ constant and $P_4=p_\phi=$ constant.

On the other hand, the coordinates $r$ and $\theta$ appear in the Hamiltonian \ref{ham} and the corresponding conjugate momenta $p_r$ and $p_\theta$ are not constants of motion. Hamilton's equations of motion in this sector of the phase space are found to be  
\begin{align}
 \dot p_r=-\frac{\partial \mathcal{H}}{\partial r}=-\frac{\pi \alpha^{\prime} P_0^2 f^{\prime}(r) }{f^2(r)}-\pi \alpha^{\prime}p_r^2f^{\prime}(r)+\frac{2 \pi \alpha^{\prime} p_{\theta}^2}{r^3}+\frac{2 \pi \alpha^{\prime}P_4^2 }{r^3\cos^2\theta}-\frac{k^2 r \sin^2\theta}{2 \pi \alpha^{\prime}},
\end{align}
\begin{align}
 \dot r = \frac{\partial \mathcal{H}}{\partial p_r}=2 \pi \alpha^{\prime} p_r  f(r),
\label{m2}
\end{align}
\begin{align}
 \dot p_\theta=-\frac{\partial \mathcal{H}}{\partial \theta}= -\frac{ 2 \pi \alpha^{\prime} P_4^2 \sin \theta }{r^2 \cos^3\theta}-\frac{ k^2 r^2 \sin\theta \cos\theta}{2  \pi \alpha^{\prime}},
\end{align}
and
\begin{align}
 \dot \theta=\frac{\partial \mathcal{H}}{\partial p_\theta}= \frac{2 \pi \alpha^{\prime} p_{\theta}}{r^2}.
 \label{m3}
\end{align}

Eliminating $\dot t$, $\dot r$, $\dot \theta$ and $\dot \phi$ in the constraint equation \ref{con7} by employing equations \ref{m1}, \ref{m4}, \ref{m2} and \ref{m3}, we obtain
\begin{align}
\pi \alpha^{\prime}\left( f p_r^2 +\frac{p_{\theta}^2}{r^2}+\frac{P_4^2}{r^2 \cos^2 \theta}-\frac{P_0^2}{f}\right)+\frac{1}{4\pi \alpha^{\prime}}k^2 r^2  \sin^2 \theta=0.
 \label{ham0}
\end{align}
Comparing \ref{ham0} with \ref{ham}, and noting that $p_t=P_0$ and $p_\phi=P_4$ are the constants of motion, we have 
\begin{align}
 \mathcal{H}=0.
 \label{cont}
\end{align}

We shall take up a canonical quantization approach to obtain the wave function of the string in the background of the Schwarzschild black hole in the following sections.

\section{Quantum dynamics of the string }
\label{qm}

The quantum mechanical equation for the wave function $\Psi$ of the system is obtained by promoting the classical Hamiltonian  \ref{ham} to an operator $\mathcal{H} \to \mathcal{\hat H}$, leading to    
\begin{align}
  \mathcal{\hat H} \Psi= 0.
 \label{weqn}
\end{align}
In the Schr\"odinger (string coordinate) representation, equation \ref{weqn} can be reexpressed as 
\begin{align}
\left[\pi \alpha^{\prime}\hbar^2\left( f   \frac{\partial^2 }{\partial r^2} +\frac{1}{r^2}  \frac{\partial^2 }{\partial \theta^2}+\frac{1}{r^2 \cos^2 \theta} \frac{\partial^2 }{\partial \phi^2}-\frac{1}{f}\frac{\partial^2 }{\partial t^2}\right)-\frac{1}{4\pi \alpha^{\prime}}  k^2 r^2 \sin^2 \theta\right]\Psi=0,
\label{waveeqn}
\end{align}
where $\Psi(r,\theta,\phi,t)$ is the wave function of the string.

The time coordinate in equation \ref{waveeqn} can be separated by writing $\Psi(r,\theta,\phi,t)=\psi(r,\theta,\phi) T(t)$, yielding the following two equations:
\begin{align}
\left[\pi \alpha^{\prime}\hbar^2\left( f^2   \frac{\partial^2 }{\partial r^2} +\frac{f}{r^2}  \frac{\partial^2 }{\partial \theta^2}+ \frac{f}{r^2 \cos^2 \theta} \frac{\partial^2 }{\partial \phi^2}\right)- \frac{k^2}{4\pi \alpha^{\prime}}  r^2 f \sin^2 \theta+\varepsilon^2\right]\psi=0
\label{tsepeqn}
\end{align}
and
\begin{align}
 \left(\frac{\partial^2 }{\partial t^2}+\frac{\varepsilon^2}{\pi \alpha^{\prime}\hbar^2}\right)T=0,
 \label{teqn}
\end{align}
where $-\varepsilon^2$ is the separation constant.

The general solution of equation \ref{teqn} is sinusoidal, given by 
\begin{align}
 T(t)=C_1 e^{i\varepsilon t/\hbar\sqrt{\pi \alpha^{\prime}}}+C_2 e^{-i\varepsilon t/\hbar\sqrt{\pi \alpha^{\prime}}},
 \label{tsol}
\end{align}
where $C_1$ and $C_2$ are arbitrary constants. 

Moreover, the coordinate $\phi$ is separable in equation \ref{tsepeqn} upon writing $\psi(r,\theta,\phi)=\mathcal{Z}(r,\theta)\Phi(\phi)$. This substitution gives rise to the following two equations:
\begin{align}
\left[\pi \alpha^{\prime}\hbar^2\left\{r^2  f(r)  \frac{\partial^2 }{\partial r^2} + \frac{\partial^2 }{\partial \theta^2}\right\}-   \frac{k^2 r^4}{4\pi \alpha^{\prime}}   \sin^2 \theta+\frac{\varepsilon^2 r^2  }{f(r)}-\frac{K^2}{\cos^2 \theta}\right]\mathcal{Z}(r,\theta)=0
\label{phisepsim}
\end{align}
and 
\begin{align}
 \left(\frac{\partial^2 }{\partial \phi^2}+\frac{K^2}{\pi \alpha^{\prime}\hbar^2}\right)\Phi(\phi)=0,
 \label{phieqn}
\end{align}
where $-K^2$ is the separation constant.

Equation \ref{phieqn} admits oscillatory solutions, the general form being
\begin{align}
 \Phi=C_3  e^{iK\phi /\hbar\sqrt{\pi \alpha^{\prime}}}+C_4 e^{-iK\phi /\hbar\sqrt{\pi \alpha^{\prime}}},
 \label{phisol}
\end{align}
where $C_3$ and $C_4$ are arbitrary constants.

It remains to solve equation \ref{phisepsim}, a partial differential equation in two variables, with the string coordinates $r$ and $\theta$ intricately coupled with each other.  Nevertheless, this is not a hindrance, as we shall see that the dynamical behavior of the string can be confined in a small radial region exterior to the black hole.

\begin{figure}
\centering
 \includegraphics{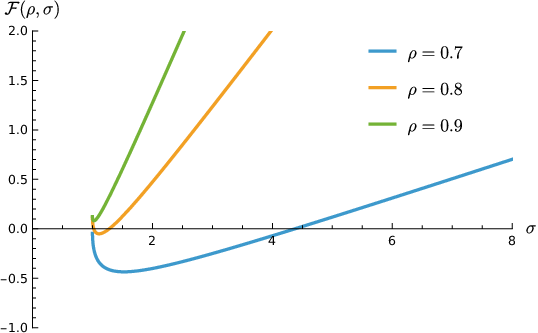}
   \caption{The function $\mathcal{F}(\rho,\sigma)$ for three values of $\rho$ consistent with the condition \ref{range}. 
For $\rho= 0.7, 0.8$, the respective values of $\mathcal{F}(\rho,\sigma)$ are positive for 
$\sigma>\sigma_*=4.37658, 1.28033$, the respective zeros of $\mathcal{F}(\rho,\sigma)$. For $\rho=0.9$, $\mathcal{F}(\rho,\sigma)$ is always positive consistent with the condition \ref{sigma}.}
 \label{rootF}
\end{figure}

\section{Dynamics at the trapping radius}
\label{trap}

As will be clear later, there exists a trapping radius in the exterior region of the AdS$_5$ black hole with the metric function 
\begin{align}
 f=1-\frac{\mu}{r^2}+\lambda r^2,
 \label{fr}
\end{align}
where $\mu$ and $\lambda$ are positive constants.

In order to obtain the dynamics of the string in the vicinity of the trapping radius, we write $r=r_0+x$, where $r_0$ is a constant and $x$ is a small radial displacement.

In equation \ref{phisepsim}, the term $r^4\sin^2\theta$ poses difficulty in separating the radial and angular coordinates. However, since we are interested in the behavior of the string in close vicinity of the trapping radius, we replace this term by $r_0^4\sin^2 \theta $, leading to 
\begin{align}
\left[\pi \alpha^{\prime}\hbar^2\left\{r^2  f(r)  \frac{\partial^2 }{\partial r^2} + \frac{\partial^2 }{\partial \theta^2}\right\}-   \frac{k^2 r_0^4}{4\pi \alpha^{\prime}}   \sin^2 \theta+\frac{\varepsilon^2 r^2  }{f(r)}-\frac{K^2}{\cos^2 \theta}\right]\mathcal{Z}(r,\theta)\approx 0
\label{nearHeqn}
\end{align}

\begin{figure}
\centering
 \includegraphics{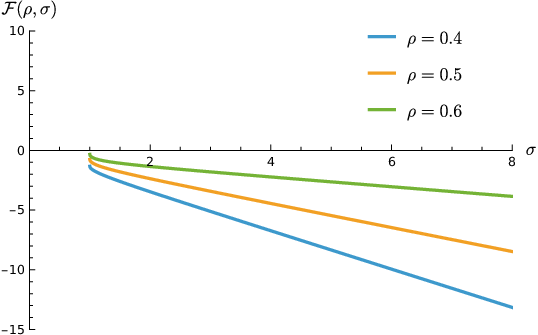}
   \caption{The function $\mathcal{F}(\rho,\sigma)$ for three values of $\rho$ violating the condition \ref{range}. 
For $\rho= 0.4, 0.5, 0.6$, the respective values of $\mathcal{F}(\rho,\sigma)$ are always negative for $\sigma>1.03822$.}
 \label{negroot}
\end{figure}

A separation of variables  can be implemented by writing   
\begin{equation}
 \mathcal{Z}(r,\theta)=\mathcal{R}(r)\Theta(\theta),
 \label{rthep1}
\end{equation}
leading to 
\begin{align}
\left[ \pi \alpha^{\prime}\hbar^2r^2  f(r) \frac{\partial^2 }{\partial r^2}+\frac{\varepsilon^2 r^2  }{f(r)}-\kappa\right]\mathcal{R}(r)=0
\label{rad}
\end{align}
 and 
\begin{align}
\left[\pi \alpha^{\prime}\hbar^2\frac{\partial^2 }{\partial \theta^2}-  \frac{k^2 r_0^4}{4\pi \alpha^{\prime}}   \sin^2\theta-\frac{K^2}{\cos^2 \theta}+\kappa\right]\Theta(\theta)=0,
\label{angu}
\end{align}
where $\kappa$ is the separation constant.

Equation \ref{angu} requires special attention and we shall solve it after solving  the $x$-equation \ref{rad}  near the trapping radius. 
Expanding about the trapping radius by writing $r=r_0+x$, we can cast equation \ref{rad} in the form 
\begin{align}
 \left[-\pi \alpha^{\prime}\hbar^2 \frac{\partial^2 }{\partial x^2}+\alpha_1+\alpha_2 x+ \alpha_3 x^2\right]\mathcal{R}(x)=0, 
 \label{WF}
\end{align}
up to $\mathcal{O}(x^2)$,  where
\begin{align}
\alpha_1= \frac{\kappa}{r_0^2 \left(1-\frac{\mu}{r_0^2}+\lambda r_0^2\right)}-\frac{\varepsilon^2}{\left(1-\frac{\mu}{r_0^2}+\lambda r_0^2\right)^2}, 
\label{a1}
\end{align}
\begin{align}
  \alpha_2= \frac{4\varepsilon^2 \left(\mu+\lambda r_0^4\right) }{r_0^3\left(1-\frac{\mu}{r_0^2}+\lambda r_0^2\right)^{3}}-\frac{2 \kappa \left(1+2 \lambda r_0^2\right)}{r_0^3\left(1-\frac{\mu}{r_0^2}+\lambda r_0^2\right)^{2}},
  \label{a2}
\end{align}
and
\begin{align}
 \alpha_3=\frac{\kappa \left(6 \mu \lambda r_0^2+\mu+10 \lambda^2 r_0^6+9 \lambda r_0^4+3 r_0^2\right)}{r_0^6\left(1-\frac{\mu}{r_0^2}+\lambda r_0^2\right)^{3}}-\frac{2 \varepsilon^2 \left(3 \mu^2+16 \mu \lambda r_0^4+3 \mu r_0^2+5 \lambda^2 r_0^8-\lambda r_0^6\right)}{r_0^6\left(11-\frac{\mu}{r_0^2}+\lambda r_0^2\right)^{4}}.
 \label{a3}
\end{align}

Equation \ref{WF} can be rewritten as 
\begin{align}
  \left[-\pi \alpha^{\prime}\hbar^2 \frac{\partial^2 }{\partial x^2} + \alpha_3\left(x+\frac{1}{2}\frac{\alpha_2}{\alpha_3}\right)^2\right]\mathcal{R}(x)=\left(\frac{1}{4}\frac{\alpha_2^2}{\alpha_3}-\alpha_1\right)\mathcal{R}(x),
  \label{rx}
\end{align}
which takes the form of standard Schr\"odinger equation for a one dimensional harmonic oscillator, given by 
\begin{align}
  \left[-\frac{\hbar^2}{2m} \frac{\partial^2 }{\partial \xi^2}+\frac{1}{2} m\omega^2\xi^2\right]\mathcal{R}(\xi)=E\mathcal{R}(\xi),
  \label{sho}
\end{align}
where $m=\frac{1}{\sqrt{\pi \alpha^{\prime}}}$, $\omega^2=\alpha_3$, $\xi=x+\frac{1}{2}\frac{\alpha_2}{\alpha_3}$ and $E=\frac{1}{2}\frac{1}{\sqrt{\pi \alpha^{\prime}}}\left(\frac{1}{4}\frac{\alpha_2^2}{\alpha_3}-\alpha_1\right)$.

The eigenvalues and the normalized eigen functions in the Schrodinger equation  \ref{sho} are given by
\begin{align}
 E_n=\left(n+\frac{1}{2}\right)\hbar\omega
\end{align}
and 
\begin{align}
 \mathcal{R}_n(\xi)=\left(\frac{a}{\sqrt{\pi}2^n n!}\right)^{1/2} e^{-\frac{1}{2}a^2\xi^2} H_n(a \xi),
 \label{Hermite}
\end{align}
where $a=\left(\frac{m\omega}{\hbar}\right)^{1/2}$, and $H_n(a \xi)$ are the standard Hermite polynomials of order $n$, with the quantum number $n=0,1,2,\ldots$ \cite{LandauQM}.

The validity of \ref{sho} requires the conditions $\omega^2>0$ and $E>0$. Since they are defined by the parameters $\alpha_1$, $\alpha_2$ and $\alpha_3$, which are in turn defined by \ref{a1}, \ref{a2} and \ref{a3} that contain two independent separation constant $\kappa$ and $\varepsilon^2$, those conditions can always be satisfied.

\begin{figure}
\centering
 \includegraphics{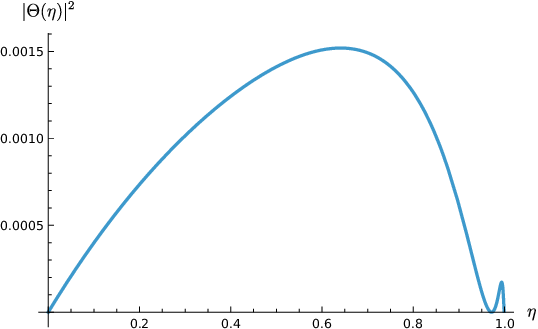}
 \caption{Angular probability density $|\Theta(\eta)|^2$ given by the solution \ref{THETA}, with $c_2=0$ in \ref{ysol}, for $\alpha=1$, $\beta=1$, $\gamma=\frac{3}{16}$, and $\delta=1$.}
 \label{thetac1}
\end{figure}

\begin{figure}
\centering
 \includegraphics{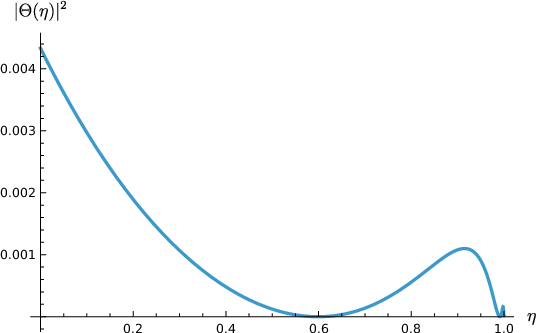}
 \caption{Angular probability density $|\Theta(\eta)|^2$ given by the solution \ref{THETA}, with $c_1=0$ in \ref{ysol}, for $\alpha=1$, $\beta=1$, $\gamma=\frac{3}{16}$, and $\delta=1$.}
 \label{thetac2}
\end{figure}

\section{Existence of Stretched Horizon}

Since $\omega^2=\alpha_3$, equation \ref{a3} suggests that the condition $\omega^2>0$ can be satisfied in appropriate domains for the values of $\kappa$ and $\varepsilon^2$. Upon choosing such appropriate domains with $\alpha_3>0$, the condition $E>0$ gives two conditions from its definition given by  $E=\frac{1}{2}\frac{1}{\sqrt{\pi \alpha^{\prime}}}\left(\frac{1}{4}\frac{\alpha_2^2}{\alpha_3}-\alpha_1\right)$. 

In case it so turns out that $\alpha_1\leq0$ according to its definition \ref{a1}, then $E>0$ is immediately guaranteed from its definition.
On the other hand, if it turns out that $\alpha_1>0$, then we require $\alpha_2^2>4\alpha_1 \alpha_3$, in order to satisfy the condition $E>0$. 
Upon meeting these conditions, equation \ref{sho} can represent a quantum mechanical harmonic oscillator in the variable $\xi$. Thus there are three cases depending on the magnitude of $\alpha_1$, that we shall discuss shortly.

The variable $\xi$ can be related to the radial coordinate as $r=r_0-\frac{1}{2}\frac{\alpha_2}{\alpha_3}+\xi$. Thus about the point $r_{\rm trap}=r_0-\frac{1}{2}\frac{\alpha_2}{\alpha_3}$, the simple harmonic oscillator can execute small oscillation with displacement $\xi$, which is described quantum mechanically by equation \ref{sho}. Effectively, the string gets trapped at the trapping radius $r_{\rm trap}$ and it makes  quantized simple harmonic radial motion about the trapping radius. The 2-sphere defined by this trapping radius is analogous to Susskind's  {\em stretched horizon}.

To establish the existence of the stretched horizon at the trapping radius, we shall consider the three cases in the following.

\begin{figure}
\centering
 \includegraphics{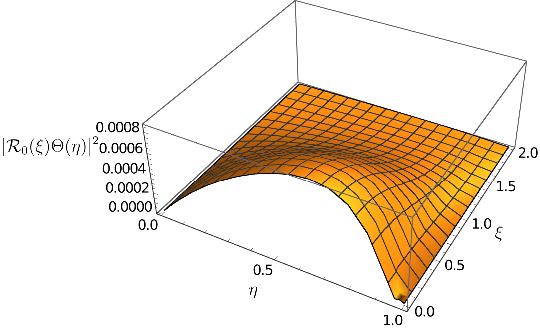}
 \caption{Combined probability density $|\mathcal{R}_n(\xi)\Theta(\eta)|^2$ given by the solutions \ref{Hermite} and \ref{THETA}, with $n=0$ and $a=1$ in \ref{Hermite} and $c_2=0$ in \ref{ysol}, for $\alpha=1$, $\beta=1$, $\gamma=\frac{3}{16}$, and $\delta=1$.}
 \label{R0c1}
\end{figure}

\subsection{Case I: $\alpha_1=0$}

We first consider the case $\alpha_1=0$. From the definition of $\alpha_1$ given by \ref{a1}, this yields $\varepsilon^2 r_0^4-\kappa r_0^2+\kappa \mu=0$, upon neglecting $\lambda$, considering the smallness of the cosmological constant. The roots of this equation are therefore found to be 
\begin{align}
 r_0^2=\frac{\kappa \pm \sqrt{\kappa(\kappa-4 \varepsilon^2 \mu)}}{2 \varepsilon^2}.
 \label{root1}
\end{align}
Since $r_0$ must be real-valued, this gives the condition $\kappa>4 \varepsilon^2 \mu$. Consequently, we write $\kappa=4 s \varepsilon^2 \mu$, with $s>1$.
Thus, we can rewrite the roots \ref{root1} as
\begin{align}
 r_0^2=2 s \mu \pm 2 \sqrt{s \mu^2(s-1)}.
 \label{rot1}
\end{align}

From the definition of $\alpha_3$ given by \ref{a3}, the requirement $\omega^2=\alpha_3>0$ leads to the inequality
\begin{align}
 3 r_0^4-5 \mu r_0^2-6 \mu^2>0,
\end{align}
which, upon using \ref{rot1} with the upper $(+)$ sign, gives the condition
\begin{align}
 24 s^2+24 s \sqrt{s (s-1)}-22 s -10 \sqrt{s (s-1)}-6>0.
 \label{ine1}
\end{align}
This condition is always satisfied for 
\begin{align}
 s>\frac{91+11 \sqrt{97}}{192} \approx 1.03822.
\end{align}

Upon meeting this condition, the Schr\"odinger equation \ref{sho} meets all requirements to describe a quantum harmonic oscillator, namely, $\omega^2=\alpha_3>0$ and $E=\frac{1}{8\sqrt{\pi\alpha'}}\frac{\alpha_2^2}{\alpha_3}>0$, implying the existence of a trapping radius giving rise to a stretched horizon in the case $\alpha_1=0$.

It may be checked that the condition $\alpha_3>0$ cannot be met if the lower $(-)$ sign is chosen in \ref{rot1}.

\begin{figure}
\centering
 \includegraphics{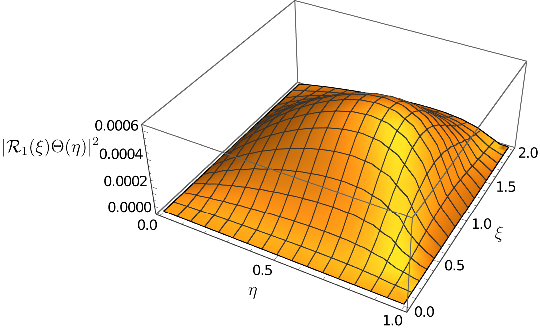}
 \caption{Combined probability density $|\mathcal{R}_n(\xi)\Theta(\eta)|^2$ given by the solutions \ref{Hermite} and \ref{THETA}, with $n=1$ and $a=1$ in \ref{Hermite} and $c_2=0$ in \ref{ysol}, for $\alpha=1$, $\beta=1$, $\gamma=\frac{3}{16}$, and $\delta=1$.}
 \label{R1c1}
\end{figure}

\subsection{Case II: $\alpha_1<0$ }

For the case $\alpha_1<0$, the definition of $\alpha_1$ given by \ref{a1} leads to the inequality $\varepsilon^2 r_0^4>\kappa (r_0^2 -\mu)$. 
Thus writing $\varepsilon^2 r_0^4=p\kappa (r_0^2 -\mu)$ with $p>1$, we have 
\begin{align}
  \varepsilon^2 r_0^4-p\kappa (r_0^2 -\mu)=0,
  \label{a3c}
\end{align}
upon neglecting $\lambda$, since the cosmological constant is very small.
The roots of equation \ref{a3c} are given by 
\begin{align}
 r_0^2=\frac{p \kappa \pm \sqrt{p \kappa(p \kappa-4 \varepsilon^2 \mu)}}{2 \varepsilon^2}.
 \label{root2}
\end{align}
For the roots to be  real, we must have $p \kappa> 4 \varepsilon^2 \mu$, or equivalently
\begin{align}
 p \kappa= 4 q \varepsilon^2 \mu,
\end{align}
with $q>1$. Inserting this expression for $p\kappa$ in \ref{root2}, the positive root is found to be 
\begin{align}
 r_0^2=2q \mu +2 \sqrt{q \mu^2(q-1)},
 \label{rot2}
\end{align}
upon choosing the upper sign in \ref{root2}.

\begin{figure}
\centering
 \includegraphics{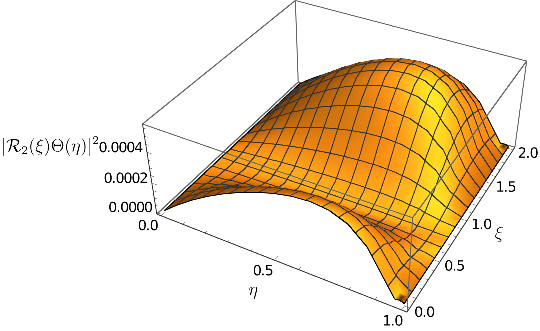}
 \caption{Combined probability density $|\mathcal{R}_n(\xi)\Theta(\eta)|^2$ given by the solutions \ref{Hermite} and \ref{THETA}, with $n=2$ and $a=1$ in \ref{Hermite} and $c_2=0$ in \ref{ysol}, for $\alpha=1$, $\beta=1$, $\gamma=\frac{3}{16}$, and $\delta=1$.}
 \label{R2c1}
\end{figure}

For the necessary condition of $\omega^2=\alpha_3>0$, we obtain from the definition of $\alpha_3$ given in \ref{a3} the inequality
\begin{align}
 3 r_0^4-5 \mu r_0^2-6 \mu^2>0.
 \label{a3d}
\end{align}

Substituting the root \ref{rot2} in \ref{a3d}, we arrive at the inequality
\begin{align}
 24 q^2+24 q\sqrt{q (q-1)}-22 q -10 \sqrt{q (q-1)}-6>0,
 \label{ine2}
\end{align}
that can always be satisfied for 
\begin{align}
q>\frac{91+11 \sqrt{97}}{192}\approx 1.03822.
\end{align}

Upon meeting this condition, the Schr\"odinger equation \ref{sho} meets all requirements to describe a quantum harmonic oscillator, namely, $\omega^2=\alpha_3>0$ and $E=\frac{1}{2\sqrt{\pi\alpha'}}\left(\frac{\alpha_2^2}{4\alpha_3}-\alpha_1\right)>0$, implying the existence of a trapping radius giving rise to a stretched horizon in the case $\alpha_1<0$.

\subsection{Case III: $\alpha_1>0$ and $\alpha_2^2>4\alpha_1 \alpha_3$}

For the case $\alpha_1>0$, the definition of $\alpha_1$ given by \ref{a1} leads to the inequality $\varepsilon^2 r_0^4<\kappa (r_0^2 -\mu)$. 
Thus writing $\varepsilon^2 r_0^4=\rho\kappa (r_0^2 -\mu)$ with $\rho<1$, we have 
\begin{align}
  \varepsilon^2 r_0^4-\rho\kappa (r_0^2 -\mu)=0
  \label{a3e}
\end{align}
upon neglecting $\lambda$, since the cosmological constant is very small.
The roots of equation \ref{a3e} are given by 
\begin{align}
 r_0^2=\frac{\rho \kappa \pm \sqrt{\rho \kappa(\rho \kappa-4 \varepsilon^2 \mu)}}{2 \varepsilon^2}
 \label{root3}
\end{align}
For the roots to be  real, we must have $\rho\kappa> 4 \varepsilon^2 \mu$, or equivalently
\begin{align}
 \rho\kappa= 4 \sigma \varepsilon^2 \mu,
 \label{kappa}
\end{align}
with $\sigma>1$. Inserting this expression for $\rho\kappa$ in \ref{root3}, the positive root is found to be 
\begin{align}
 r_0^2=2\sigma \mu +2 \sqrt{\sigma \mu^2(\sigma-1)},
 \label{rot3}
\end{align}
upon choosing the upper sign in \ref{root3}.

For the necessary condition of $\omega^2=\alpha_3>0$, we obtain from the definition of $\alpha_3$ given in \ref{a3} the inequality
\begin{align}
 3 r_0^4-5 \mu r_0^2-6 \mu^2>0.
 \label{a3f}
\end{align}

\begin{figure}
\centering
 \includegraphics{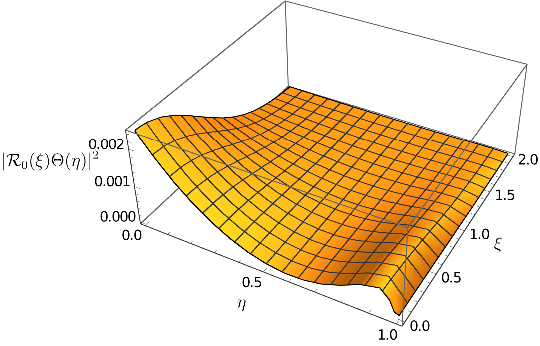}
 \caption{Combined probability density $|\mathcal{R}_n(\xi)\Theta(\eta)|^2$ given by the solutions \ref{Hermite} and \ref{THETA}, with $n=0$ and $a=1$ in \ref{Hermite} and $c_1=0$ in \ref{ysol}, for $\alpha=1$, $\beta=1$, $\gamma=\frac{3}{16}$, and $\delta=1$.}
 \label{R0c2}
\end{figure}

Substituting the root \ref{rot3} in \ref{a3f}, we arrive at the inequality
\begin{align}
 24 \sigma^2+24 \sigma \sqrt{\sigma (\sigma-1)}-22 \sigma -10 \sqrt{\sigma (\sigma-1)}-6>0,
 \label{ine3}
\end{align}
that can always be satisfied for 
\begin{align}
\sigma>\frac{91+11 \sqrt{97}}{192} \approx 1.03822.
\label{sigma}
\end{align}

Apart from the positivity of $\omega^2=\alpha_3$, we require in this case of $\alpha_1>0$, an additional condition 
\begin{align}
 \alpha_2^2>4\alpha_1 \alpha_3
\end{align}
to meet the requirement $E=\frac{1}{2\sqrt{\pi\alpha'}}\left(\frac{\alpha_2^2}{4\alpha_3}-\alpha_1\right)>0$.

Using the definitions of $\alpha_1$, $\alpha_2$, and $\alpha_3$ from \ref{a1}, \ref{a2}, and \ref{a3}, and eliminating $\kappa$ by substituting from \ref{kappa}, we arrive at the inequality
\begin{align}
 6 \mu ^2 r_0^4 \sigma  (\rho +4 \sigma )+3 \rho  r_0^8 (2 \sigma-\rho  )-8 \mu ^4 \sigma ^2-12 \mu ^3 \rho  r_0^2 \sigma -\mu  r_0^6 (\rho ^2+16 \sigma ^2)>0.
 \label{a3g}
\end{align}
Substituting the root \ref{rot3} in \ref{a3g}, we find the inequality 
\begin{align}
\mathcal{F}(\rho,\sigma)=\rho(15 \sigma +\rho-8 \rho\sigma-3)+(8\rho-1)(\rho-1)\sqrt{\sigma(\sigma-1)}-5\sigma>0.
\label{fcon}
\end{align}

The condition $\mathcal{F}(\rho,\sigma)>0$ can be satisfied only for $\rho>\frac{2}{3}$ along with $\sigma>\sigma_*$ where $\sigma_*$ is the highest root of  $\mathcal{F}(\rho,\sigma)=0$, with the value of $\rho$ in the range
\begin{align}
 \frac{2}{3}<\rho<1.
 \label{range}
\end{align}

Figure \ref{rootF} displays the function $\mathcal{F}(\rho,\sigma)$ for three values of $\rho$ consistent with the condition \ref{range}. 
For the values of $\rho= 0.7, 0.8$, the plots show that the respective values of $\mathcal{F}(\rho,\sigma)$ are positive for 
$\sigma>\sigma_*=4.37658, 1.28033$, which are the respective zeros of $\mathcal{F}(\rho,\sigma)$. For $\rho=0.9$, the function $\mathcal{F}(\rho,\sigma)$ is always positive (no negative part) consistent with the previous condition 
$\sigma>1.03822$, given by \ref{sigma}.

On the other hand, Figure \ref{negroot} shows the function $\mathcal{F}(\rho,\sigma)$ for three values of $\rho$ violating the condition \ref{range}. 
For the values of $\rho= 0.4, 0.5, 0.6$, the plots show that the respective values of $\mathcal{F}(\rho,\sigma)$ are always negative for 
all values of $\sigma>1.03822$.

Upon meeting the above conditions, the Schr\"odinger equation \ref{sho} meets all requirements to describe a quantum harmonic oscillator, namely, $\omega^2=\alpha_3>0$ and $E=\frac{1}{2\sqrt{\pi\alpha'}}\left(\frac{\alpha_2^2}{4\alpha_3}-\alpha_1\right)>0$, implying the existence of a trapping radius giving rise to a stretched horizon in the case $\alpha_1>0$ and $\alpha_2^2>4\alpha_1\alpha_3$.

\begin{figure}
\centering
 \includegraphics{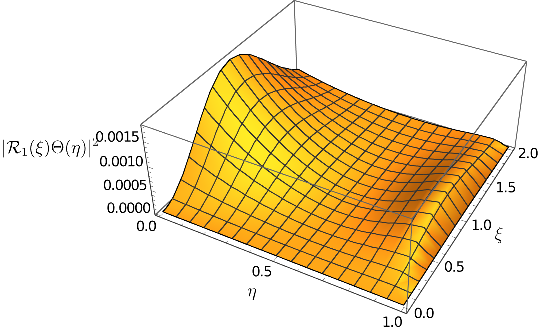}
 \caption{Combined probability density $|\mathcal{R}_n(\xi)\Theta(\eta)|^2$ given by the solutions \ref{Hermite} and \ref{THETA}, with $n=1$ and $a=1$ in \ref{Hermite} and $c_1=0$ in \ref{ysol}, for $\alpha=1$, $\beta=1$, $\gamma=\frac{3}{16}$, and $\delta=1$.}
 \label{R1c2}
\end{figure}

\begin{figure}
\centering
 \includegraphics{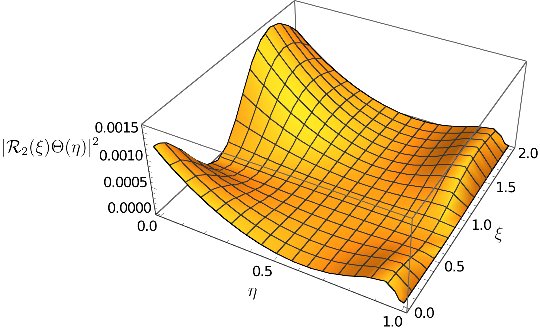}
 \caption{Combined probability density $|\mathcal{R}_n(\xi)\Theta(\eta)|^2$ given by the solutions \ref{Hermite} and \ref{THETA}, with $n=1$ and $a=1$ in \ref{Hermite} and $c_1=0$ in \ref{ysol}, for $\alpha=1$, $\beta=1$, $\gamma=\frac{3}{16}$, and $\delta=1$.}
 \label{R2c2}
\end{figure}

\section{Angular Sector}
\label{ang}

The angular part \ref{angu} can be  cast into the form 
\begin{align}
\left[\frac{\partial^2 }{\partial \theta^2} -k_1 \sin^2 \theta-\frac{k_2}{\cos^2 \theta}+k_3\right]\Theta(\theta)=0,
\label{thedeqn}
\end{align}
where 
\begin{align}
k_1=\frac{1}{\pi \alpha^{\prime}\hbar^2}  \frac{r_0^4 k^2}{4\pi \alpha^{\prime}}, 
\hskip0.5cm k_2=\frac{K^2}{\pi \alpha^{\prime}\hbar^2}\hskip0.25cm {\rm and} \hskip0.25cm
k_3=\frac{\kappa}{\pi \alpha^{\prime}\hbar^2}.
\end{align}

Defining $\eta=\sin^2\theta$, equation \ref{thedeqn} transforms into
\begin{align}
\left[4\eta(1-\eta)\frac{\partial^2 }{\partial \eta^2}+2\left(1-2\eta\right)\frac{\partial }{\partial \eta} -k_1 \eta-\frac{k_2}{1-\eta}+k_3\right]\Theta(\eta)=0, 
\label{yeqn}
\end{align}
which can further be rewritten as 
\begin{align}
\left[\frac{\partial^2 \Theta}{\partial \eta^2}+\frac{1}{2}\left(\frac{1}{\eta}+\frac{1}{\eta-1}\right)\frac{\partial \Theta}{\partial \eta} +\frac{A}{\eta}+\frac{B}{\eta-1}+\frac{C}{(\eta-1)^2}\right]\Theta(\eta)=0,
\label{yeqnps}
\end{align}
where $4A=k_3-k_2$, $4B=k_1-k_2-k_3$ and $4C=-k_2$.

Defining $\mathcal{Y}(\eta)=[\eta(\eta-1)]^{1/4} \Theta(\eta)$, equation \ref{yeqnps} can be simplified to 
\begin{align}
\frac{\partial^2 \mathcal{Y}}{\partial \eta^2}+\left[\frac{\alpha}{\eta}+\frac{\beta}{(\eta-1)}+\frac{\gamma}{\eta^2}+\frac{\delta}{(\eta-1)^2}\right]\mathcal{Y}(\eta)=0,
\label{fzeqn}
\end{align}
where $\alpha=A+\frac{1}{8}$, $\beta=B-\frac{1}{8}$, $\gamma=\frac{3}{16}$ and $\delta=C+\frac{3}{16}$.

Equation \ref{fzeqn} is the confluent Heun equation whose general solution is 
%\begin{widetext}
\begin{align}
\nonumber
 \mathcal{Y}(\eta)=& c_1 \eta^{\frac{1}{2}+\frac{1}{2} \sqrt{1-4 \gamma }} (\eta-1)^{\frac{1}{2}+\frac{1}{2} \sqrt{1-4 \delta }}\\
\nonumber & \times {\rm HeunC}\left[\alpha -\textstyle\frac{1}{2} (1+\sqrt{1-4 \gamma }) (1+\sqrt{1-4 \delta }),\alpha +\beta ,1+\sqrt{1-4 \gamma },1+\sqrt{1-4 \delta },0,\eta\right]\\
\nonumber & +c_2 \eta^{\frac{1}{2}-\frac{1}{2} \sqrt{1-4 \gamma }} (\eta-1)^{\frac{1}{2}+\frac{1}{2} \sqrt{1-4 \delta }}\\
& \times {\rm HeunC}\left[\alpha -\textstyle\frac{1}{2} (1-\sqrt{1-4 \gamma }) (1+\sqrt{1-4 \delta }),\alpha +\beta ,1-\sqrt{1-4 \gamma },1+\sqrt{1-4 \delta },0,\eta\right],
 \label{ysol}
\end{align}
%\end{widetext}
where HeunC[$a_1,a_2,a_3,a_4,a_5,\eta$] is the  confluent Heun function and $c_1$, $c_2$ are arbitrary constant.

Thus, the  general solution of the angular equation \ref{thedeqn} can be obtained from 
\begin{align}
 \Theta(\eta)=[\eta(\eta-1)]^{-1/4}\mathcal{Y}(\eta)
 \label{THETA}
\end{align}
upon using the general solution \ref{ysol}.

Figure \ref{thetac1} illustrates the angular probability density $|\Theta(\eta)|^2$ given by the solution \ref{THETA}, with $c_2=0$ in \ref{ysol}, for $\alpha=1$, $\beta=1$, $\gamma=\frac{3}{16}$, and $\delta=1$. 

Similarly, Figure \ref{thetac2} shows the angular probability density $|\Theta(\eta)|^2$ given by the solution \ref{THETA}, with $c_1=0$ in \ref{ysol}, for $\alpha=1$, $\beta=1$, $\gamma=\frac{3}{16}$, and $\delta=1$.

Figures \ref{R0c1}, \ref{R1c1}, \ref{R2c1} show the combined probability density $|\mathcal{R}_n(\xi)\Theta(\eta)|^2$ given by the solutions \ref{Hermite} and \ref{THETA}, with $n=0, 1, 2$ (respectively) and $a=1$ in \ref{Hermite} and $c_2=0$ in \ref{ysol}, for $\alpha=1$, $\beta=1$, $\gamma=\frac{3}{16}$, and $\delta=1$.

Furthermore, Figures \ref{R0c2}, \ref{R1c2}, \ref{R2c2} show the combined probability density $|\mathcal{R}_n(\xi)\Theta(\eta)|^2$ given by the solutions \ref{Hermite} and \ref{THETA}, with $n=0, 1, 2$ (respectively) and $a=1$ in \ref{Hermite} and $c_1=0$ in \ref{ysol}, for $\alpha=1$, $\beta=1$, $\gamma=\frac{3}{16}$, and $\delta=1$.

\section{Discussion and Conclusion}

\label{concl}

In this work, we explored the dynamics of a closed bosonic string in the curved spacetime of an AdS$_5$ Schwarzschild black hole.  Describing the string with the Polyakov action, we obtained the Lagrangian of the string in the AdS$_5$ background with a suitable embedding. This facilitated a smooth transition to Hamiltonian description. Applying the canonical quantization scheme, we obtained a Schr\"odinger-type equation that governs the quantum string behavior in the radial and angular coordinates.

A key result of our analysis is the identification of a ``trapping radius'' in the exterior region of the black hole, which has profound implications for the string dynamics. This trapping radius indicates that the string is confined so that its radial motion is like a quantum harmonic oscillator with discrete energy levels. 
 
Moreover, we showed that the angular part of the string’s equation of motion is governed by the confluent Heun equation, which is a second-order linear differential equation.  The confluent Heun function as a solution to the angular sector indicates that the string's angular motion exhibits intricate behavior.

The two-dimensional spherical surface defined by the trapping radius exhibits striking similarities to ``stretched horizon'' introduced by Susskind in the context of black hole complementarity. Similarly to Susskind’s framework, the trapping radius identified in our analysis functions as a confining surface where the closed string becomes localized, undergoing quantized oscillatory motion. This localization may reflect a quantum mechanical description of the black hole’s microscopic degrees of freedom, suggesting that the trapping radius may play a role analogous to the stretched horizon in mediating interaction between gravity and string dynamics. In this sense, the string’s behavior at the trapping radius may offer a window into the quantum structure of spacetime.

In conclusion, our novel study provides a comprehensive analysis of the dynamics of a closed bosonic string in the AdS$_5$ Schwarzschild black hole background, revealing the emergence of a trapping radius, quantized oscillatory motion, and intricate angular dynamics. The quantized oscillations of the string near the trapping surface endow it with properties closely resembling those of a Planckian black-body radiator. This analogy arises naturally from the fact that the string, when localized near the trapping radius by an effective harmonic potential, exhibits discrete energy levels corresponding to its vibrational modes. These quantized excitations, when thermally populated, can lead to a spectrum of radiation that mimics black-body emission at a characteristic temperature associated with the black hole horizon.

This observation offers a compelling microscopic perspective on the origin of Hawking radiation, suggesting that the thermal spectrum traditionally derived via semiclassical quantum field theory in curved spacetime may, in fact, have an underlying string interpretation. In particular, it points toward a mechanism whereby the thermal emission from black holes can be understood as resulting from the quantized vibrations of strings that are dynamically confined to a finite region just outside the event horizon.

\section*{Acknowledgments}
Harpreet Singh is grateful to Prof.~Gordon L.~Kane of the University of Michigan for motivating discussions. Harpreet Singh is supported by the Prime Minister's Research Fellowship (PMRF ID: 1902180), Ministry of Education, Government of India.

%\bibliographystyle{unsrturl}
%\bibliography{SBH.bib}

\begin{thebibliography}{10}

\bibitem{PhysRevD.93.084012}
Arman Tursunov, Zden\ifmmode \check{e}\else~\v{e}\fi{}k Stuchl\'{\i}k, and
  Martin Kolo\ifmmode~\check{s}\else \v{s}\fi{}.
\newblock Circular orbits and related quasiharmonic oscillatory motion of
  charged particles around weakly magnetized rotating black holes.
\newblock {\em Phys. Rev. D}, 93:084012, Apr 2016.
\newblock URL: \url{https://link.aps.org/doi/10.1103/PhysRevD.93.084012}, \href
  {https://doi.org/10.1103/PhysRevD.93.084012}
  {\path{doi:10.1103/PhysRevD.93.084012}}.

\bibitem{PhysRevD.83.124015}
Tim Johannsen and Dimitrios Psaltis.
\newblock Metric for rapidly spinning black holes suitable for strong-field
  tests of the no-hair theorem.
\newblock {\em Phys. Rev. D}, 83:124015, Jun 2011.
\newblock URL: \url{https://link.aps.org/doi/10.1103/PhysRevD.83.124015}, \href
  {https://doi.org/10.1103/PhysRevD.83.124015}
  {\path{doi:10.1103/PhysRevD.83.124015}}.

\bibitem{Panis2019}
Radim P{\'a}nis, Martin Kolo{\v{s}}, and Zden{\v{e}}k Stuchl{\'i}k.
\newblock Determination of chaotic behaviour in time series generated by
  charged particle motion around magnetized schwarzschild black holes.
\newblock {\em The European Physical Journal C}, 79(6):479, Jun 2019.
\newblock \href {https://doi.org/10.1140/epjc/s10052-019-6961-7}
  {\path{doi:10.1140/epjc/s10052-019-6961-7}}.

\bibitem{PhysRevD.95.084037}
Bobir Toshmatov, Zden\ifmmode \check{e}\else~\v{e}\fi{}k Stuchl\'{\i}k, and
  Bobomurat Ahmedov.
\newblock Generic rotating regular black holes in general relativity coupled to
  nonlinear electrodynamics.
\newblock {\em Phys. Rev. D}, 95:084037, Apr 2017.
\newblock URL: \url{https://link.aps.org/doi/10.1103/PhysRevD.95.084037}, \href
  {https://doi.org/10.1103/PhysRevD.95.084037}
  {\path{doi:10.1103/PhysRevD.95.084037}}.

\bibitem{PhysRevD.83.044053}
Ahmadjon Abdujabbarov, Bobomurat Ahmedov, and Abdullo Hakimov.
\newblock Particle motion around black hole in ho\ifmmode \check{r}\else
  \v{r}\fi{}ava-lifshitz gravity.
\newblock {\em Phys. Rev. D}, 83:044053, Feb 2011.
\newblock URL: \url{https://link.aps.org/doi/10.1103/PhysRevD.83.044053}, \href
  {https://doi.org/10.1103/PhysRevD.83.044053}
  {\path{doi:10.1103/PhysRevD.83.044053}}.

\bibitem{PhysRevD.99.044012}
Carlos~A. Benavides-Gallego, Ahmadjon Abdujabbarov, Daniele Malafarina,
  Bobomurat Ahmedov, and Cosimo Bambi.
\newblock Charged particle motion and electromagnetic field in $\gamma$
  spacetime.
\newblock {\em Phys. Rev. D}, 99:044012, Feb 2019.
\newblock URL: \url{https://link.aps.org/doi/10.1103/PhysRevD.99.044012}, \href
  {https://doi.org/10.1103/PhysRevD.99.044012}
  {\path{doi:10.1103/PhysRevD.99.044012}}.

\bibitem{Yi_2020}
Miao Yi and Xin Wu.
\newblock Dynamics of charged particles around a magnetically deformed
  schwarzschild black hole.
\newblock {\em Physica Scripta}, 95(8):085008, jul 2020.
\newblock URL: \url{https://dx.doi.org/10.1088/1402-4896/aba4c2}, \href
  {https://doi.org/10.1088/1402-4896/aba4c2}
  {\path{doi:10.1088/1402-4896/aba4c2}}.

\bibitem{PhysRevD.50.R618}
C.~P. Dettmann, N.~E. Frankel, and N.~J. Cornish.
\newblock Fractal basins and chaotic trajectories in multi-black-hole
  spacetimes.
\newblock {\em Phys. Rev. D}, 50:R618--R621, Jul 1994.
\newblock URL: \url{https://link.aps.org/doi/10.1103/PhysRevD.50.R618}, \href
  {https://doi.org/10.1103/PhysRevD.50.R618}
  {\path{doi:10.1103/PhysRevD.50.R618}}.

\bibitem{doi:10.1142/S0217732307022815}
William Hanan and Eugen Radu.
\newblock Chaotic motion in multi-black hole spacetimes and holographic
  screens.
\newblock {\em Modern Physics Letters A}, 22(06):399--406, 2007.
\newblock \href
  {http://arxiv.org/abs/https://doi.org/10.1142/S0217732307022815}
  {\path{arXiv:https://doi.org/10.1142/S0217732307022815}}, \href
  {https://doi.org/10.1142/S0217732307022815}
  {\path{doi:10.1142/S0217732307022815}}.

\bibitem{Hosur2016}
Pavan Hosur, Xiao-Liang Qi, Daniel~A. Roberts, and Beni Yoshida.
\newblock Chaos in quantum channels.
\newblock {\em Journal of High Energy Physics}, 2016(2):4, Feb 2016.
\newblock \href {https://doi.org/10.1007/JHEP02(2016)004}
  {\path{doi:10.1007/JHEP02(2016)004}}.

\bibitem{Gur-Ari2016}
Guy Gur-Ari, Masanori Hanada, and Stephen~H. Shenker.
\newblock Chaos in classical d0-brane mechanics.
\newblock {\em Journal of High Energy Physics}, 2016(2):91, Feb 2016.
\newblock \href {https://doi.org/10.1007/JHEP02(2016)091}
  {\path{doi:10.1007/JHEP02(2016)091}}.

\bibitem{PhysRevD.95.066014}
Pallab Basu, Pankaj Chaturvedi, and Prasant Samantray.
\newblock Chaotic dynamics of strings in charged black hole backgrounds.
\newblock {\em Phys. Rev. D}, 95:066014, Mar 2017.
\newblock URL: \url{https://link.aps.org/doi/10.1103/PhysRevD.95.066014}, \href
  {https://doi.org/10.1103/PhysRevD.95.066014}
  {\path{doi:10.1103/PhysRevD.95.066014}}.

\bibitem{PhysRevD.89.086011}
Da-Zhu Ma, Jian-Pin Wu, and Jifang Zhang.
\newblock Chaos from the ring string in a gauss-bonnet black hole in ads$_5$
  space.
\newblock {\em Phys. Rev. D}, 89:086011, Apr 2014.
\newblock URL: \url{https://link.aps.org/doi/10.1103/PhysRevD.89.086011}, \href
  {https://doi.org/10.1103/PhysRevD.89.086011}
  {\path{doi:10.1103/PhysRevD.89.086011}}.

\bibitem{Ma2020}
Da-Zhu Ma, Dan Zhang, Guoyang Fu, and Jian-Pin Wu.
\newblock Chaotic dynamics of string around charged black brane with
  hyperscaling violation.
\newblock {\em Journal of High Energy Physics}, 2020(1):103, Jan 2020.
\newblock \href {https://doi.org/10.1007/JHEP01(2020)103}
  {\path{doi:10.1007/JHEP01(2020)103}}.

\bibitem{Ma2022}
Da-Zhu Ma, Fang Xia, Dan Zhang, Guo-Yang Fu, and Jian-Pin Wu.
\newblock Chaotic dynamics of string around the conformal black hole.
\newblock {\em The European Physical Journal C}, 82(4):372, Apr 2022.
\newblock \href {https://doi.org/10.1140/epjc/s10052-022-10338-5}
  {\path{doi:10.1140/epjc/s10052-022-10338-5}}.

\bibitem{Maldacena2016}
Juan Maldacena, Stephen~H. Shenker, and Douglas Stanford.
\newblock A bound on chaos.
\newblock {\em Journal of High Energy Physics}, 2016(8):106, Aug 2016.
\newblock \href {https://doi.org/10.1007/JHEP08(2016)106}
  {\path{doi:10.1007/JHEP08(2016)106}}.

\bibitem{PandoZayas2010}
Leopoldo~A. Pando~Zayas and C{\'e}sar~A. Terrero-Escalante.
\newblock Chaos in the gauge/gravity correspondence.
\newblock {\em Journal of High Energy Physics}, 2010(9):94, Sep 2010.
\newblock \href {https://doi.org/10.1007/JHEP09(2010)094}
  {\path{doi:10.1007/JHEP09(2010)094}}.

\bibitem{LI2025139164}
Kai Li, Da-Zhu Ma, and Zhen-Meng Xu.
\newblock Chaotic dynamics of string around the charged kiselev black hole.
\newblock {\em Physics Letters B}, 860:139164, 2025.
\newblock URL:
  \url{https://www.sciencedirect.com/science/article/pii/S0370269324007226},
  \href {https://doi.org/https://doi.org/10.1016/j.physletb.2024.139164}
  {\path{doi:https://doi.org/10.1016/j.physletb.2024.139164}}.

\bibitem{DEVEGA1987320}
H.J. {De Vega} and N.~Sanchez.
\newblock A new approach to string quantization in curved spacetimes.
\newblock {\em Physics Letters B}, 197(3):320--326, 1987.
\newblock URL:
  \url{https://www.sciencedirect.com/science/article/pii/0370269387903923},
  \href {https://doi.org/https://doi.org/10.1016/0370-2693(87)90392-3}
  {\path{doi:https://doi.org/10.1016/0370-2693(87)90392-3}}.

\bibitem{PhysRevD.51.6917}
H.~J. de~Vega, A.~L. Larsen, and N.~S\'anchez.
\newblock Semiclassical quantization of circular strings in de sitter and
  anti--de sitter spacetimes.
\newblock {\em Phys. Rev. D}, 51:6917--6928, Jun 1995.
\newblock URL: \url{https://link.aps.org/doi/10.1103/PhysRevD.51.6917}, \href
  {https://doi.org/10.1103/PhysRevD.51.6917}
  {\path{doi:10.1103/PhysRevD.51.6917}}.

\bibitem{Ishii2015}
Takaaki Ishii and Keiju Murata.
\newblock Turbulent strings in ads/cft.
\newblock {\em Journal of High Energy Physics}, 2015(6):86, Jun 2015.
\newblock \href {https://doi.org/10.1007/JHEP06(2015)086}
  {\path{doi:10.1007/JHEP06(2015)086}}.

\bibitem{Ishii2016}
Takaaki Ishii and Keiju Murata.
\newblock Dynamical ads strings across horizons.
\newblock {\em Journal of High Energy Physics}, 2016(3):35, Mar 2016.
\newblock \href {https://doi.org/10.1007/JHEP03(2016)035}
  {\path{doi:10.1007/JHEP03(2016)035}}.

\bibitem{PhysRevD.98.086007}
Koji Hashimoto, Keiju Murata, and Norihiro Tanahashi.
\newblock Chaos of wilson loop from string motion near black hole horizon.
\newblock {\em Phys. Rev. D}, 98:086007, Oct 2018.
\newblock URL: \url{https://link.aps.org/doi/10.1103/PhysRevD.98.086007}, \href
  {https://doi.org/10.1103/PhysRevD.98.086007}
  {\path{doi:10.1103/PhysRevD.98.086007}}.

\bibitem{Maldacena1999}
Juan Maldacena.
\newblock The large-n limit of superconformal field theories and supergravity.
\newblock {\em International Journal of Theoretical Physics}, 38(4):1113--1133,
  Apr 1999.
\newblock \href {https://doi.org/10.1023/A:1026654312961}
  {\path{doi:10.1023/A:1026654312961}}.

\bibitem{PhysRevD.53.3296}
H.~J. de~Vega and I.~L. Egusquiza.
\newblock Strings next to and inside black holes.
\newblock {\em Phys. Rev. D}, 53:3296--3307, Mar 1996.
\newblock URL: \url{https://link.aps.org/doi/10.1103/PhysRevD.53.3296}, \href
  {https://doi.org/10.1103/PhysRevD.53.3296}
  {\path{doi:10.1103/PhysRevD.53.3296}}.

\bibitem{FROLOV1989255}
V.P. Frolov, V.D. Skarzhinsky, A.I. Zelnikov, and O.~Heinrich.
\newblock Equilibrium configurations of a cosmic string near a rotating black
  hole.
\newblock {\em Physics Letters B}, 224(3):255--258, 1989.
\newblock URL:
  \url{https://www.sciencedirect.com/science/article/pii/0370269389912252},
  \href {https://doi.org/https://doi.org/10.1016/0370-2693(89)91225-2}
  {\path{doi:https://doi.org/10.1016/0370-2693(89)91225-2}}.

\bibitem{PhysRevD.47.4498}
C.~O. Loust\'o and N.~S\'anchez.
\newblock String instabilities in black hole spacetimes.
\newblock {\em Phys. Rev. D}, 47:4498--4509, May 1993.
\newblock URL: \url{https://link.aps.org/doi/10.1103/PhysRevD.47.4498}, \href
  {https://doi.org/10.1103/PhysRevD.47.4498}
  {\path{doi:10.1103/PhysRevD.47.4498}}.

\bibitem{PhysRevD.49.763}
H.~J. de~Vega and I.~L. Egusquiza.
\newblock Strings in cosmological and black hole backgrounds: Ring solutions.
\newblock {\em Phys. Rev. D}, 49:763--778, Jan 1994.
\newblock URL: \url{https://link.aps.org/doi/10.1103/PhysRevD.49.763}, \href
  {https://doi.org/10.1103/PhysRevD.49.763}
  {\path{doi:10.1103/PhysRevD.49.763}}.

\bibitem{PhysRevD.51.6929}
A.~L. Larsen and N.~S\'anchez.
\newblock New classes of exact multistring solutions in curved spacetimes.
\newblock {\em Phys. Rev. D}, 51:6929--6948, Jun 1995.
\newblock URL: \url{https://link.aps.org/doi/10.1103/PhysRevD.51.6929}, \href
  {https://doi.org/10.1103/PhysRevD.51.6929}
  {\path{doi:10.1103/PhysRevD.51.6929}}.

\bibitem{PhysRevD.50.2623}
A.~L. Larsen.
\newblock Circular string instabilities in curved spacetime.
\newblock {\em Phys. Rev. D}, 50:2623--2630, Aug 1994.
\newblock URL: \url{https://link.aps.org/doi/10.1103/PhysRevD.50.2623}, \href
  {https://doi.org/10.1103/PhysRevD.50.2623}
  {\path{doi:10.1103/PhysRevD.50.2623}}.

\bibitem{membranebook}
Kip~S. {Thorne}, Richard~H. {Price}, and Douglas~A. {MacDonald}.
\newblock {\em Black holes: The membrane paradigm}.
\newblock Yale University Press, New Haven, CT, 1986.

\bibitem{THOOFT1985727}
Gerard {'t Hooft}.
\newblock On the quantum structure of a black hole.
\newblock {\em Nuclear Physics B}, 256:727--745, 1985.
\newblock URL:
  \url{https://www.sciencedirect.com/science/article/pii/0550321385904183},
  \href {https://doi.org/https://doi.org/10.1016/0550-3213(85)90418-3}
  {\path{doi:https://doi.org/10.1016/0550-3213(85)90418-3}}.

\bibitem{THOOFT1990138}
G.~{'t Hooft}.
\newblock The black hole interpretation of string theory.
\newblock {\em Nuclear Physics B}, 335(1):138--154, 1990.
\newblock URL:
  \url{https://www.sciencedirect.com/science/article/pii/055032139090174C},
  \href {https://doi.org/https://doi.org/10.1016/0550-3213(90)90174-C}
  {\path{doi:https://doi.org/10.1016/0550-3213(90)90174-C}}.

\bibitem{GtHooft_1991}
G~'t~Hooft.
\newblock The black hole horizon as a quantum surface.
\newblock {\em Physica Scripta}, 1991(T36):247, jan 1991.
\newblock URL: \url{https://dx.doi.org/10.1088/0031-8949/1991/T36/026}, \href
  {https://doi.org/10.1088/0031-8949/1991/T36/026}
  {\path{doi:10.1088/0031-8949/1991/T36/026}}.

\bibitem{PhysRevLett.71.2367}
Leonard Susskind.
\newblock String theory and the principle of black hole complementarity.
\newblock {\em Phys. Rev. Lett.}, 71:2367--2368, Oct 1993.
\newblock URL: \url{https://link.aps.org/doi/10.1103/PhysRevLett.71.2367},
  \href {https://doi.org/10.1103/PhysRevLett.71.2367}
  {\path{doi:10.1103/PhysRevLett.71.2367}}.

\bibitem{PhysRevD.48.3743}
Leonard Susskind, L\'arus Thorlacius, and John Uglum.
\newblock The stretched horizon and black hole complementarity.
\newblock {\em Phys. Rev. D}, 48:3743--3761, Oct 1993.
\newblock URL: \url{https://link.aps.org/doi/10.1103/PhysRevD.48.3743}, \href
  {https://doi.org/10.1103/PhysRevD.48.3743}
  {\path{doi:10.1103/PhysRevD.48.3743}}.

\bibitem{PhysRevD.49.6606}
Leonard Susskind.
\newblock Strings, black holes, and lorentz contraction.
\newblock {\em Phys. Rev. D}, 49:6606--6611, Jun 1994.
\newblock URL: \url{https://link.aps.org/doi/10.1103/PhysRevD.49.6606}, \href
  {https://doi.org/10.1103/PhysRevD.49.6606}
  {\path{doi:10.1103/PhysRevD.49.6606}}.

\bibitem{PhysRevD.50.2725}
Arthur Mezhlumian, Amanda Peet, and L\'arus Thorlacius.
\newblock String thermalization at a black hole horizon.
\newblock {\em Phys. Rev. D}, 50:2725--2730, Aug 1994.
\newblock URL: \url{https://link.aps.org/doi/10.1103/PhysRevD.50.2725}, \href
  {https://doi.org/10.1103/PhysRevD.50.2725}
  {\path{doi:10.1103/PhysRevD.50.2725}}.

\bibitem{PhysRevD.60.024012}
A.~L. Larsen and A.~Nicolaidis.
\newblock String spreading on a black hole horizon.
\newblock {\em Phys. Rev. D}, 60:024012, Jun 1999.
\newblock URL: \url{https://link.aps.org/doi/10.1103/PhysRevD.60.024012}, \href
  {https://doi.org/10.1103/PhysRevD.60.024012}
  {\path{doi:10.1103/PhysRevD.60.024012}}.

\bibitem{PhysRevD.51.1793}
David~A. Lowe and Andrew Strominger.
\newblock Strings near a rindler or black hole horizon.
\newblock {\em Phys. Rev. D}, 51:1793--1799, Feb 1995.
\newblock URL: \url{https://link.aps.org/doi/10.1103/PhysRevD.51.1793}, \href
  {https://doi.org/10.1103/PhysRevD.51.1793}
  {\path{doi:10.1103/PhysRevD.51.1793}}.

\bibitem{ALarsen_1993}
A~L Larsen.
\newblock Dynamics of cosmic strings and springs; a covariant formulation.
\newblock {\em Classical and Quantum Gravity}, 10(8):1541, aug 1993.
\newblock URL: \url{https://dx.doi.org/10.1088/0264-9381/10/8/014}, \href
  {https://doi.org/10.1088/0264-9381/10/8/014}
  {\path{doi:10.1088/0264-9381/10/8/014}}.

\bibitem{LARSEN1991375}
A.L. Larsen.
\newblock Cosmic string winding around a black hole.
\newblock {\em Physics Letters B}, 273(4):375--379, 1991.
\newblock URL:
  \url{https://www.sciencedirect.com/science/article/pii/037026939190286Y},
  \href {https://doi.org/https://doi.org/10.1016/0370-2693(91)90286-Y}
  {\path{doi:https://doi.org/10.1016/0370-2693(91)90286-Y}}.

\bibitem{ALarsen_1994}
A~L Larsen.
\newblock Chaotic string-capture by black hole.
\newblock {\em Classical and Quantum Gravity}, 11(5):1201, may 1994.
\newblock URL: \url{https://dx.doi.org/10.1088/0264-9381/11/5/008}, \href
  {https://doi.org/10.1088/0264-9381/11/5/008}
  {\path{doi:10.1088/0264-9381/11/5/008}}.

\bibitem{AndreiVFrolov_1999}
Andrei~V Frolov and Arne~L Larsen.
\newblock Chaotic scattering and capture of strings by a black hole.
\newblock {\em Classical and Quantum Gravity}, 16(11):3717, nov 1999.
\newblock URL: \url{https://dx.doi.org/10.1088/0264-9381/16/11/316}, \href
  {https://doi.org/10.1088/0264-9381/16/11/316}
  {\path{doi:10.1088/0264-9381/16/11/316}}.

\bibitem{PhysRevD.79.065029}
Ted Jacobson and Thomas~P. Sotiriou.
\newblock String dynamics and ejection along the axis of a spinning black hole.
\newblock {\em Phys. Rev. D}, 79:065029, Mar 2009.
\newblock URL: \url{https://link.aps.org/doi/10.1103/PhysRevD.79.065029}, \href
  {https://doi.org/10.1103/PhysRevD.79.065029}
  {\path{doi:10.1103/PhysRevD.79.065029}}.

\bibitem{POLYAKOV1981207}
A.M. Polyakov.
\newblock Quantum geometry of bosonic strings.
\newblock {\em Physics Letters B}, 103(3):207--210, 1981.
\newblock URL:
  \url{https://www.sciencedirect.com/science/article/pii/0370269381907437},
  \href {https://doi.org/https://doi.org/10.1016/0370-2693(81)90743-7}
  {\path{doi:https://doi.org/10.1016/0370-2693(81)90743-7}}.

\bibitem{Green_Schwarz_Witten_2012}
Michael~B. Green, John~H. Schwarz, and Edward Witten.
\newblock {\em Superstring Theory}.
\newblock Cambridge Monographs on Mathematical Physics. Cambridge University
  Press, 2012.

\bibitem{Polchinski_1998}
Joseph Polchinski.
\newblock {\em String Theory}.
\newblock Cambridge Monographs on Mathematical Physics. Cambridge University
  Press, 1998.

\bibitem{Zwiebach_2009}
Barton Zwiebach.
\newblock {\em A First Course in String Theory}.
\newblock Cambridge University Press, 2 edition, 2009.

\bibitem{Becker_Becker_Schwarz_2006}
Katrin Becker, Melanie Becker, and John~H. Schwarz.
\newblock {\em String Theory and M-Theory: A Modern Introduction}.
\newblock Cambridge University Press, 2006.

\bibitem{LandauQM}
Lev~Davidovich Landau and E.~M. Lifshits.
\newblock {\em Quantum Mechanics}, volume~3 of {\em Course of Theoretical
  Physics}.
\newblock Butterworth-Heinemann, Oxford, 1991.
\newblock \href {https://doi.org/10.1016/C2013-0-02793-4}
  {\path{doi:10.1016/C2013-0-02793-4}}.

\end{thebibliography}

\end{document}